\def \ba {\begin{array}}
\def \ea {\end{array}}

\def \bea {\begin{eqnarray}}
\def \eea {\end{eqnarray}}

\def \be {\begin{equation}}
\def \ee {\end{equation}}

\def\ni{\noindent}
\def\nn{\nonumber}

\def\l[{\left[}
\def\r]{\right]}

\documentclass[11pt,a4]{article}
\usepackage{amssymb,amsfonts}
\begin{document}

\begin{center} 
{\bf Extended diffeomorphism algebras in (quantum) gravitational physics}
\footnote{Work partially
supported by the DGICYT.}
\end{center}
\bigskip
\bigskip
\centerline{ {\it V. Aldaya\footnote{E-mail: valdaya@iaa.es} 
and J.L. Jaramillo\footnote{E-mail: jarama@iaa.es}} }
\bigskip

\begin{itemize}
\item {Instituto de Astrof\'{\i}sica de Andaluc\'{\i}a (CSIC), Apartado Postal
 3004, 18080 Granada, Spain.}
\item  {Instituto Carlos I de F\'\i sica Te\'orica y Computacional, Facultad
de Ciencias, Universidad de Granada, Campus de Fuentenueva, 
Granada 18002, Spain.} 
\end{itemize}

\bigskip
\begin{center}
{\bf Abstract}
\end{center}
\small

\begin{list}{}{\setlength{\leftmargin}{3pc}\setlength{\rightmargin}{3pc}}
\item  We construct an explicit representation of the algebra of
local diffeomorphisms of a manifold with realistic dimensions. This is
achieved in the setting of a general approach to the (quantum) dynamics
of a physical system which is characterized by the 
fundamental role assigned to a basic underlying symmetry.
The developed mathematical formalism makes contact with the relevant
gravitational notions by means of the addition of some extra structure.
The specific manners in which this is accomplished, together with their
corresponding physical interpretation, lead to different 
gravitational models. Distinct strategies 
are in fact briefly outlined, showing the versatility of the present 
conceptual framework. 
\end{list}

\normalsize

\vskip 1cm

\section{Introduction}

A model for space-time dynamics in 
$1+1-$dimensions was introduced in \cite{QG1199}. The present 
paper studies the necessary extension to realistic dimensions 
of the involved mathematical structures  
together with their physical implications.

The original work was conceived as a directly quantum formulation
of the physical system, rather than a {\it quantization} of a previously 
devised classical theory. The rationale for such a construction was 
the fundamental role attached to the
notion of symmetry in the characterization of the dynamics.
Specifically, group 
cohomology governing the structure of the central extensions became the
guiding principle in the identification of those dynamical degrees of freedom
building the phase space. 
The choice of the abstract Virasoro algebra as the basic symmetry, led
to a quantum theory making contact with two-dimensional Polyakov gravity 
\cite{Polyakov} at the semi-classical level.
The main achievements of that particular model were
the recovery of a space-time notion out of the symmetry itself
without its explicit introduction into scheme from the very beginning, 
on the one hand, and the
gain of dynamical character by some diffeomorphisms thus entailing a quantum 
breakdown of their associated gauge invariance, on the other hand.

When extending this kind of analysis to systems with support on
higher-dimensional space-times, we can identify two conceptually different
levels of generalization. The first one, structural 
in nature, deals with the construction of the mathematical formalism
supporting the new model. We shall address this problem by firstly 
identifying a 
Lie algebra which properly generalizes the Virasoro one, and then constructing
the quantum and semi-classical dynamics that it provides. This task will
be achieved by using the so-called Group Approach to Quantization (GAQ)
\cite{GAQ}, as was the case in \cite{QG1199}.
The second level refers to the physical interpretation of that
mathematical formalism. The intuition provided in the two-dimensional case
by the Polyakov action will be definitively lost in the higher-dimensional
model, and therefore new links with a {\it metric} notion  will 
be needed. Our specific goal at this level will be the implementation of
the quantum breakdown of diffeomorphism invariance at the quantum level, 
perhaps the most fundamental feature 
in this approach. With regard to the space-time recovery from the group
structure we shall only dwell on the most simple situation in which no
{\it critical} conditions in the cohomology parameters (i.e. anomalies)
occur. Adopting this
attitude, we are essentially leaving this specific issue open for 
future research.  

\medskip 

Apart from the above-described line of reasoning, the present considerations
make an interesting contact with the research program initiated in 
\cite{electrograv}. A novel approach to the problem of mixing between
space-time and internal gauge interactions was proposed
with support on a {\it revisited gauge principle} which implies a non-trivial
interplay between the corresponding local current algebras. The restriction 
in these works to a Particle-Mechanical setting freezes the
{\it dynamics} corresponding to those  local currents, therefore hiding their
relevance. 
Nevertheless, these intrinsic dynamics prove to be an {\it essential}
ingredient in the GAQ treatment of gauge interactions \cite{Manolo1,Manolo2} 
at the Field level (see below).
In particular, gravitational interaction is addressed in \cite{electrograv}
by making local 
the translation subgroup \cite{Kibble} which leads to a current algebra
isomorphic to that of local diffeomorphisms (see subsection 2.2). 
Therefore the dynamics
tied to this diffeomorphism algebra provides an intermediate step 
fundamental in the understanding of the proposed interaction mixing for 
field degrees of freedom, constituting a secondary but significant 
goal of the present paper. 

\medskip

The structure of the paper is as follows.  Section 2 is divided in two
subsections. First one briefly presents the fundamentals of the GAQ, the
technique we use to construct the 
dynamical system out of a Lie symmetry. In the second one, the rigour of the
presentation is significatly  
relaxed in order to motivate the relevant algebra we are using as a starting
point. It also includes a brief explanation on the 
role of local gauge groups in the GAQ treatment of gauge theories.
Section 3 recovers a more formal style and includes  our main results, 
i.e.,  the dynamics associated with the Lie algebra
proposed in the previous section. This will be done both at the 
semi-classical and quantum level. In particular, a maximum-weight 
representation of an (non-centrally) extended
diffeomorphism algebra will be explicitly constructed. Section 4
contains the corresponding physical discussion and, finally, the conclusions
will be presented in section 5.

\section{General setting}

\subsection{Technical formalism: GAQ}

We briefly describe the basics of the Group Approach to 
Quantization (GAQ), a formalism that surpasses its role as a technichal 
device, 
deeply influencing and guiding the conceptual understanding of the
physical dynamics construction.

The starting point is a Lie group $\tilde{G}$ with a $U(1)$-principal 
bundle structure with base $G$. This bundle possesses a connection 1-form 
$\Theta$ which generalizes the Poincar\'e-Cartan form $\Theta_{PC}$, i.e. the
(pre-)contact form of time-dependent dynamics. This connection
also plays a main role in the quantization procedure, and is (naturally) 
selected 
among the components of the left-invariant, Lie-algebra valued, Maurer-Cartan
1-form, in such a way that $\Theta(\Xi)=1$ and 
$\Theta(\tilde{X}^L_i)=0$, where $\Xi$ is the vertical (or fundamental)
vector field of the principal bundle and $\tilde{X}^L_i$ are the rest of the 
independent left-invariant generators in a basis of the Lie algebra.

Taking $\tilde{G}$ as a central extension of $G$ by $U(1)$
(the {\it tilde} will denote {\it extended objects} throughout) we
consider the space of complex functions $\Psi$ on $\tilde{G}$
satisfying the $U(1)$-equivariance condition, i.e. $\Psi(\zeta*g)=
\zeta\Psi(g)$, $\zeta\in U(1), g\in \tilde{G}$. On this space, the left 
translations on $\tilde{G}$
(generated by right-invariant vector fields $\tilde{X}^R$) realize a 
representation of the group which is in general reducible. Right translations,
generated by left-invariant vector fields $\tilde{X}^L$, 
do commute with the left ones on any Lie group. Therefore, certain subgroups,
called {\it polarization subgroups } $G_{\cal P}$, can be chosen to perform
a reduction of the representation through the so-called {\it polarization 
conditions}, following the language of Geometric Quantization \cite{Woodhouse},
which are implemented by left-invariant vector fields.
This defines a reduced subspace of wave functions ${\cal H}$ satisfying 
$\Psi(g*G_{\cal P})=\Psi(g)$  (or infinitesimally $L_{\tilde{X}^L}\Psi=0$,
$\forall\tilde{X}^L\in{\cal G_P}$, where ${\cal G_P}$ is the Lie algebra of 
$G_{\cal P}$).

In the finite-dimensional case, the Hilbert space structure on the space of 
wave functions is provided
by the invariant Haar-like measure constructed from the exterior product of the
left-invariant canonical form components, 
$\Omega^L\equiv\theta^{Lg^1}\wedge\theta^{Lg^2}\wedge...$ On the reduced
space ${\cal H}$ (at least for the case of first-order polarizations), 
the existence of a quasi-invariant measure is granted, since the wave 
functions have support on $G/G_{\cal P}$, which is a homogeneous space. 
However we shall deal here with an infinite-dimensional symmetry whose 
associated representation space is given by an orbit of the group through
a (unique) {\it vacuum} state. This allows the introduction of a 
Hilbert product by 
fixing the normalization of that vacuum and imposing 
a rule of {\it adjointness} for the basic operators. 

The classical theory for the system is easily recovered by defining the
Noether invariants as ${\cal N}_{g^i}\equiv i_{\tilde{X}^R_{g^i}}\Theta$. 
They are in fact invariant under the evolution generated by vector fields
inside ${\cal G}_\Theta\equiv Ker\Theta\cap Ker d\Theta$, which define the 
generalized equations of motion. The classical phase space is given 
by $G/G_{\Theta}$ (see \cite{GAQ} and references therein). 
A Poisson bracket can be introduced (defined by $d\Theta$) in such a way 
that the Noether invariants generate a Lie algebra isomorphic to that
of $\tilde{G}$.

\subsection{Starting algebra}

The choice of the starting Lie algebra, defining the fundamental
symmetry underlying the system, will be guided by the attempt to implement our 
basic goals.

\vskip 0.6cm

\ni {\it Generalization of Virasoro algebra}

\medskip

\ni Firstly, the generalization of the Virasoro algebra, together with the
desire to analyse a potential physical breakdown of the diffeomorphism 
invariance
in realistic space-time dimensions, suggests the adoption of the local 
diffeomorphism algebra of a $d-$dimensional manifold $M$ ($d\geq 2$) as a 
natural departure point. This algebra, denoted by $Vect(M)$ in \cite{Fuks},
is simply given by vector fields on $M$ with the natural Lie product,
\bea
[{\cal L}_\eta, {\cal L}_\xi]={\cal L}_{[\eta,\xi]} \label{Lieder}
\eea
where $[\eta,\xi]=(\eta^\mu \partial_\mu\xi^\nu-\xi^\mu \partial_\mu\eta^\nu)
\partial_\nu$. 
The analysis of the associated dynamics is more easily developed by 
fixing a basis for the vector fields. When choosing $L_\mu({\bf m})=
ie^{-im_\rho x^\rho}\partial_\mu$, with ${\bf m}$ a $d-$dimensional vector
of integer entries, (\ref{Lieder}) is written,
\bea
[L_\mu({\bf m}),L_\nu({\bf n})]=n_\mu L_\nu({\bf m+n})-
m_\nu L_\mu({\bf m+n}) \ \ .
\eea
The cohomological analysis of this algebra rules out the possibility of 
non-trivial central extensions ($d\geq 2$), which is not
a fundamental drawback for the construction of the corresponding dynamics,
as will be explicitly shown at the end of this section. Non-central
extensions must be considered if one wants to enrich the dynamics, entailing
the introduction of new fields in the model.
A particular class of these extensions, the tensorial ones, are 
studied and classified in  \cite{Larsson}, \cite{Dhu} providing a variety 
of possibilities from which to choose. Here we shall consider the simplest 
extension, represented by
a generator $S^\rho({\bf m})$. This option constitutes in fact a 
direct generalization of the cubic cocycle
of the Virasoro algebra, reducing to it in the one-dimensional case (to 
be precise, a closure condition must be imposed to $S^\rho({\bf m})$).  
The Lie algebra is given by:
\bea
[L_\mu({\bf m}), L_\nu({\bf n})]&=&n_\mu L_\nu({\bf m+n})-m_\nu 
L_\mu({\bf m+n})- m_\mu n_\nu(m_\rho-n_\rho)S^\rho({\bf m+n}) 
\label{diffext} \\
\left[L_\mu({\bf m}),S^\nu({\bf n})\right]&=&n_\mu S^\nu({\bf m+n})+
\delta^\nu_\mu m_\rho S^\rho({\bf m+n}) \;\;.\nn
\eea

\medskip

\ni{\it Contact with the interaction mixing problem}

\medskip

\ni Secondly, and in order to address our purpose of relating this algebra
with the works in \cite{electrograv}, let us briefly sketch a comparison
between the ways in which field degrees of freedom are handled in
the standard approach to (Yang-Mills) gauge theories and in GAQ.

In an oversimplified version of the former, which omits all technical 
and conceptual subtleties, 
the field degrees of freedom are constructed in terms
of the components $A^a_\mu$ ({\it gauge fields}) of a connection 
on a principal bundle, with
$U$ as the structure group and the space-time $M$ as its base. These modes 
couple to the matter degrees
of freedom, living on an associated vector bundle, by means of a the so-called 
{\it minimal coupling}, technically implemented by replacing the partial 
derivatives in the matter Lagrangian with a covariant derivative defined in 
terms of the connection (very roughly, $\partial_\mu\rightarrow
(\partial_\mu+A_\mu)$). The Lagrangian giving the intrinsic dynamics of 
the {\it gauge fields} is defined as a function of the curvature 
$F_{\mu\nu}$ of the connection form \cite{Utiyama}. This description of the
system is overcomplete since not all the field degrees $A^a_\mu$ are in 
fact physical. Two connections $A'_\mu$ and $A_\mu$  are physically 
equivalent if they are
related by the action of the gauge group
(locally identifiable with $Map(M,U)$), $A'_\mu=u^{-1}A_\mu u+u^{-1}du$.
Taking quotient by the orbits of the gauge group, we obtained the reduced
or physical {\it phase} space.

The whole spirit and structural description changes in the GAQ.
Instead of a space of 
field degrees of freedom parametrized by $A_\mu$, on which a Lagrangian $L$
is defined and which must be reduced according to the action of a given
gauge group, in GAQ all these elements must be merged into a unique 
centrally-extended
group $\tilde{G}$. The dynamics, as well as the identification of the
physical degrees of freedom, must follow from the algorithm described
in the previous subsection. There is no obvious way of endowing the
connection components $A^a_\mu$ with a group law structure,  
closing, in addition, a 
Lie symmetry with the {\it gauge} group $Map(M,U)$. An explicit solution
to this problem, not necessarily unique, is presented in 
\cite{Manolo1,Manolo2}.
The price we must pay for closing such a unified Lie structure is the loss of a
clean distinction at the group law level between the degrees of 
freedom corresponding to the {\it gauge fields} (the ``connections'') and the 
parameters corresponding to the {\it gauge group}. These modes are 
intertwined in a subtle manner in such a way that a transfer of 
dynamical content occurs between them. In order to fix ideas we comment on 
the simplest
case of a local $U(1)$ gauge group in the unified treatment of the
electromagnetic and Proca fields \cite{Manolo1}.
Working in momentum space and parametrizing the gauge fields by $a_\mu(k)$ and
$a^\dagger_\mu(k)$, and the $U(1)$ parameters by $\phi(k)$ and $\phi^+(k)$, the
corresponding group law implies at the Lie algebra level the 
following commutators,
\bea
\left[\tilde{X}^L_{a^\dagger_\mu(k)}, \tilde{X}^L_{a_\mu(k')}\right]&=&
i\eta^{\mu\nu}\Delta_{kk'}\Xi \nn \\
\left[\tilde{X}^L_{\phi^\dagger(k)}, \tilde{X}^L_{\phi(k')}\right]&=&
ik^2\Delta_{kk'}\Xi  \\
\left[\tilde{X}^L_{a_\mu(k)}, \tilde{X}^L_{\phi^\dagger(k')}\right]&=&
k^\mu\Delta_{kk'}\Xi \nn \\
\left[\tilde{X}^L_{a^\dagger_\mu(k)}, \tilde{X}^L_{\phi(k')}\right]&=&
k^\mu\Delta_{kk'}\Xi \nn \ \ ,
\eea
where $\Delta_{kk'}=2k^0\delta^3(k-k')$ is the generalized delta function
on the positive sheet of the mass hyperboloid.
The second commutator explicitly shows the non-empty symplectic content
of $\phi(k)$, whereas the third and forth implies
the above-mentioned mixing between the $a_\mu(k)$ and $\phi(k)$ variables.
The diagonalization of the cocycle when determining the canonical
symplectic variables
entails the transfer of dynamical content between modes.
The generators playing the role of authentic gauge tranformations in this
setting (they
leave the reduced physical phase space invariant pointwise) are in fact,
\bea
\tilde{X}^L_{c(k)}&=&\tilde{X}^L_{\phi(k)}+ik_\mu\tilde{X}^L_{a_\mu(k)} \\
\tilde{X}^L_{c^\dagger(k)}&=&\tilde{X}^L_{\phi^\dagger(k)}+ik_\mu
\tilde{X}^L_{a^\dagger_\mu(k)}\nn
\eea
which imply the equivalence condition,
\bea
a_\mu(k)\sim a_\mu(k)+ik_\mu c(k) \ \ , \ \ \phi(k)\sim \phi(k)+c(k)
\eea
together with its hermitian conjugate, where $c(k)$ is the evolution 
parameter of the
$\tilde{X}^L_{c(k)}$ vector field.  They simply represent the momentum 
version of the standard gauge transformations. 

The only aim in the precedent discussion was that of underlying the 
importance of a good understanding of the (abstract) intrinsic dynamics
associated with the local gauge group, 
determined by its extensions, when one is addressing gauge
theories in the GAQ formalism. 

\ni Coming back to the idea of mixing external and internal interactions
as described in \cite{electrograv}, the relevant gauge invariance 
groups were determined from the local current algebras 
${\cal F}(M)\otimes{\cal G}$, where ${\cal F}(M)$ is the real functions 
on $M$ and 
${\cal G}$ the Lie algebra of the rigid part of the gauge group. Choosing
a realization $\{X_a\}$ for the latter, we have:
\bea
[f\otimes X_a,g\otimes X_b]=(f L_{X_a}g)\otimes X_b-(gL_{X_b}f)\otimes X_a+
(fg)\otimes[X_a,X_b] \label{current} \ \ .
\eea
The gravitational interaction was addressed by making local the
rigid translation group. Choosing $\partial_\mu$ as a realization for its
generators, the corresponding current algebra turns out to be isomorphic 
to the local diffeomorphism algebra (\ref{Lieder}). Therefore, in the
study of Gravity as a gauge interaction with a gauge group obtained
by making local the space-time translations (\cite{Kibble} and first part
of IIIA in \cite{Al89}), the analysis of the 
intrinsic dynamics derived from the
diffeomorphism algebra plays a fundamental role in a GAQ setting.
The modes $L_\mu({\bf m})$ are just the analogue to $\phi(k)$ in the 
electromagnetic case.
Of course, it represents only an intermediate step, since the 
{\it gauge fields}
analogue to $a_\mu(k)$\footnote{In principle several possibilities arise
ranging from metric variables to connection ones in the spirit of Ashtekar
variables.} should be consistenly introduced at a second stage.

The mixing of gravitational and electromagnetic interactions in 
\cite{electrograv} occurred through the 
weaving together the corresponding current algebras. In order to 
contemplate such a possibility, we introduce the new generators $f\otimes \Xi$ 
corresponding to the
local $U(1)$. In the momentum space we are 
dealing with, their expression is $A({\bf m})$ and the new commutators
are given by
\bea
\left[L_\mu({\bf m}),A({\bf n})\right]&=&n_\mu A({\bf m+n}) \label{A}\\
\left[A({\bf m}), A({\bf n})\right]&=&(n_\rho-m_\rho)
S^\rho({\bf m+n}) \nn \ \ ,
\eea
where we have taken advantage of the possibility of introducing a new
(and independent) extension in the $U(1)$ sector implemented by the already 
present $S^\rho({\bf m})$ generators. Therefore, our starting algebra is 
given by commutators
(\ref{diffext}) $+$ (\ref{A}).

\vskip 0.6cm

\ni{\it Central extensions}

\medskip

\ni Since we need a $U(1)$-centrally extended group in 
order to employ  the GAQ, the starting Lie algebra 
should present central terms in the commutators. The central 
terms that cannot be eliminated by a linear change of basis in the Lie algebra
are controlled at the group level \footnote{In this general remark we 
shall not address the differences between Lie
algebra and Lie group cohomology.} by the non-trivial two-cocycles
in the second cohomology 
group of the non-extended Lie group $G$, $H^2(G, U(1))$. But these are not the
only cocycles capable of creating dynamics. In fact, and it is specially 
important in 
the case of Lie groups with trivial cohomology, those coboundaries
$\xi_\lambda$ generated by a function on the group $\lambda$ such that
$\frac{\partial\lambda(g)}{\partial g^i}|_{g=e}\neq 0$ lead to
dynamics on phases spaces which are symplectomorphic to the 
coadjoint orbits of the group. Of course, these 
coboundaries generate central terms at the algebra level that could be
eliminated by linear changes, so they are consistently trivial from the
Lie structure point of view. Nevertheless,  they have crucial consequences on 
the polarization conditions and therefore, at the
representation level. We can reverse the reasoning and consider that, 
once the algebra has been (trivially or not) extended by a central 
$U(1)$ group, a linear 
change in the algebra involving  the central generator will be associated 
with one of those coboundaries creating dynamics.
These subclasses inside $H^2(G, U(1))$ were first introduced in 
\cite{Saletan} associated with some coboundaries $\xi_\lambda$, which become
non-trivial cocycles $\xi_c$ after an In\"on\"u-Wigner contraction takes 
place, and fully studied in connection with coadjoint orbits in
\cite{pseudoext}. The simplest example of the latter is 
the Poincar\'e group whose
{\it pseudo-cohomology} goes to the true cohomology of the Galilei group 
\cite{Inonu-Poin}. For finite-dimensional semi-simple groups, pseudo-cohomology
is also related to the Cech cohomology of the generalized Hopf fibrations by
the Cartan subgroups $H$, $G\rightarrow G/H$ \cite{Pepe2}.

The algebra (\ref{diffext}) $+$ (\ref{A}) we are using as the starting point 
does not present 
the central extension structure we need in order to derive a dynamics. In 
fact, no non-trivial extensions exists when the space-time dimension is over 
one. Therefore, we take advantage of the above-discussed 
{\it pseudo-cohomology} and we add a new central 
generator  $\Xi$ to the Lie algebra in a direct-sum way. We then perform 
the linear change
\bea
A({\bf m})&\mapsto& A({\bf m})+\alpha \delta({\bf m})\Xi \nn
 \\
S^\rho({\bf m})&\mapsto& S^\rho({\bf m})+C^\rho 
\delta({\bf m})\Xi  \label{central}\\
L_\mu({\bf m})&\mapsto& L_\mu({\bf m})+\tilde{C}_\mu 
\delta({\bf m})\Xi \nn \ \ ,
\eea
which will give rise to one of the dynamically non-trivial coboundaries, 
after exponentiation.

\section{Dynamics}

As we have seen, the starting point for the whole construction is the explicit
group law of the Lie symmetry defining the physical system. In the 
previous section we only
worked at the algebra level. Then, the first step now is to exponentiate it
to a finite group law. This entails the complications
derived from  dealing with infinite-dimensional Lie algebras.
We shall face this difficulty by using a perturbative order-by-order 
construction of a formal group law \cite{Serre}, as described in \cite{Pepe}.

Using an index $\mu({\bf m})$ for $L_\mu({\bf m})$, $\tilde{\mu}({\bf m})$
for $S^\mu({\bf m})$, ${\bf m}$ for $A({\bf m})$ and $\varphi$ for
$\Xi$, the structure constants in this basis are (there is a harmless global 
redefinition by a minus sign related to the particular 
exponentiation technique we are going to employ):
\bea
C^{\rho({\bf r})}_{\mu({\bf m})\nu({\bf n})}&=& -({\bf n}_\mu\delta^\rho_\nu-{\bf m}_\nu\delta^\rho_\mu)
\delta({\bf m+n-r}) \nn \\
C^{\tilde{\rho}({\bf r})}_{\mu({\bf m})\nu({\bf n})}&=& m_\mu n_\nu(m_\rho-n_\rho)
\delta({\bf m+n-r}) \nn \\
C^\varphi_{\mu({\bf m})\nu({\bf n})}&=& -\left[-m_\mu n_\nu(m_\rho-n_\rho)
C^\rho+(n_\mu\delta^\rho_\nu-m_\nu\delta^\rho_\mu)\tilde{C}_\rho\right]
\delta({\bf m+n}) \nn \\
C^{\tilde{\rho}({\bf r})}_{\mu({\bf m})\tilde{\nu}({\bf n})}&=& -(n_\mu\delta_\rho^\nu
-m_\nu\delta_\rho^\mu)\delta({\bf m+n-r}) \nn \\
C^\varphi_{\mu({\bf m})\tilde{\nu}({\bf n})}&=& -(n_\mu\delta_\rho^\nu
-m_\nu\delta_\rho^\mu)C^\rho\delta({\bf m+n})  \\
C^{\tilde{\rho}({\bf r})}_{{\bf m}{\bf n}}&=&-(n_\rho-m_\rho)\delta({\bf m+n-r}) \nn \\
C^\varphi_{{\bf m}{\bf n}}&=&-(n_\rho-m_\rho)c_2 C^\rho\delta({\bf m+n}) \nn \\
C^{\bf r}_{\mu({\bf m}){\bf n}}&=& -n_\mu\delta({\bf m+n-r}) \nn \\
C^\varphi_{\mu({\bf m}){\bf n}}&=& -\alpha n_\mu\delta({\bf m+n}) \nn \ \ ,
\eea
We use them to construct a formal group law, $g''=g'*g$, up to the third-order 
in the group variables. 
We associate
the variable $l^\mu({\bf m})$ with the generator $L_\mu({\bf m})$,
$s^\mu({\bf m})$ with $S_\mu({\bf m})$, $a({\bf m})$ with $A({\bf m})$
and $\varphi$ with $\Xi$, obtaining:
\bea
l''^\rho({\bf r})&=&l'^\rho({\bf r})+l^\rho({\bf r})+
\frac{1}{2}C^{\rho({\bf r})}_{\mu({\bf m})\nu({\bf n})}l'^\mu({\bf m})
l^\nu({\bf n})\nn\\
&+&\frac{1}{8}C^{\rho({\bf r})}_{\nu({\bf n})\mu({\bf m})}
C^{\mu({\bf m})}_{\gamma({\bf q})\sigma({\bf s})})l'^\nu({\bf n})
l'^\gamma({\bf q})l^\sigma({\bf s})+...\nn\\
s''^\rho({\bf r})&=&s'^\rho({\bf r})+s^\rho({\bf r})+
\frac{1}{2}C^{\tilde{\rho}({\bf r})}_{\mu({\bf m})\nu({\bf n})}l'^\mu({\bf m})
l^\nu({\bf n}) \nn \\
&+& C^{\tilde{\rho}({\bf r})}_{\mu({\bf m})\tilde{\nu}({\bf n})}
l'^\mu({\bf m})s^\nu({\bf n})+\frac{1}{2}
C^{\tilde{\rho}({\bf r})}_{\bf mn}a'({\bf m})a({\bf n})\nn \\
&+&\frac{1}{2}\left[\left(\frac{1}{4}C^{\tilde{\rho}({\bf r})}
_{\nu({\bf n})\mu({\bf m})}C^{\mu({\bf m})}
_{\gamma({\bf q})\sigma({\bf s})}+\frac{1}{2}
C^{\tilde{\rho}({\bf r})}
_{\nu({\bf n})\tilde{\mu}({\bf m})}C^{\tilde{\mu}({\bf m})}
_{\gamma({\bf q})\sigma({\bf s})}\right)l'^\nu({\bf n})
l'^\gamma({\bf q})l^\sigma({\bf s})\nn \right.\\
&+&\left.C^{\tilde{\rho}({\bf r})}
_{\nu({\bf n})\tilde{\mu}({\bf m})}C^{\tilde{\mu}({\bf m})}
_{\gamma({\bf q})\tilde{\sigma}({\bf s})}l'^\nu({\bf n})
l'^\gamma({\bf q})s^\sigma({\bf s})+
\frac{1}{2}C^{\tilde{\rho}({\bf r})}
_{\bf nm}C^{\bf m}_{\gamma({\bf q}){\bf s}}l'^\gamma({\bf q})
a'({\bf n})a({\bf s})\right]+... \\
a''({\bf p})&=&a'({\bf p})+a({\bf p})+C^{\bf p}_{\mu({\bf m}){\bf n}}
l'^\mu({\bf m})a({\bf n})+
\frac{1}{2}C^{\bf p}_{\nu({\bf n}){\bf m}}C^{\bf m}_{\gamma({\bf q}){\bf s}}
l'^\nu({\bf n})l'^\gamma({\bf q})a({\bf s})+...\nn \\
\varphi''&=&\varphi'+\varphi+\frac{1}{2}C^\varphi_{\mu({\bf m})\nu({\bf n})}
l'^\mu({\bf m})l^\nu({\bf n})+C^\varphi_{\mu({\bf m})\tilde{\nu}({\bf m})}
l'^\mu({\bf m})s^\nu({\bf n})+
C^\varphi_{\mu({\bf m}){\bf n}}l'^\mu({\bf m})a({\bf n}) \nn \\
&+&\frac{1}{2}C^\varphi_{\bf mn}a'({\bf m})a({\bf n})+
\frac{1}{2}\left[\left(\frac{1}{4}C^\varphi_{\rho({\bf r})\mu({\bf m})}
C^{\mu({\bf m})}_{\nu({\bf n})\sigma({\bf s})}+\frac{1}{2}
C^\varphi_{\rho({\bf r})\tilde{\mu}({\bf m})}
C^{\tilde{\mu}({\bf m})}_{\nu({\bf n})\sigma({\bf s})}\right)
l'^\rho({\bf r})l'^\nu({\bf n})l^\sigma({\bf s})\nn\right. \\
&+&C^\varphi_{\rho({\bf r})\tilde{\mu}({\bf m})}
C^{\tilde{\mu}({\bf m})}_{\nu({\bf n})\tilde{\sigma}({\bf s})}
l'^\rho({\bf r})l'^\nu({\bf n})s^\sigma({\bf s})+
C^\varphi_{\rho({\bf r}){\bf m}}
C^{\bf m}_{\nu({\bf n}){\bf s}}l'^\rho({\bf r})l'^\nu({\bf n})a({\bf s})\nn \\
&+&\left.\frac{1}{2}C^\varphi_{\bf rm}C^{\bf m}_{\nu({\bf n}){\bf s}}\right]
l'^\nu({\bf n})a'({\bf r})a({\bf s})+... \nn \ \ .
\eea
Using this, we compute the left-invariant vector fields, 
$\tilde{X}^L_{(k)}(g)=\frac{\partial g''^l(g',g)}{\partial g^{(k)}}|_
{g=e,g'=g}
\frac{\partial}{\partial g^l}$, yielding,
\bea
\tilde{X}^L_{l^\tau({\bf t})}&=&\frac{\partial}{\partial l^\tau({\bf t})}+
\left\{\frac{1}{2}C^{\rho({\bf r})}_{\mu({\bf m})\tau({\bf t})}l^\mu({\bf m})
+\frac{1}{8}C^{\rho({\bf r})}_{\nu({\bf n})\mu({\bf m})}C^{\mu({\bf m})}
_{\gamma({\bf q})\tau({\bf t})}l^\nu({\bf n})l^\gamma({\bf q})
+...\right\}\frac{\partial}{\partial l^\rho({\bf r})} \nn \\
&+&\left\{\frac{1}{2}C^{\tilde{\rho}({\bf r})}_{\mu({\bf m})\tau({\bf t})}
l^\mu({\bf m})+\left(\frac{1}{8}C^{\tilde{\rho}({\bf r})}_{\nu({\bf n})\mu({\bf m})}C^{\mu({\bf m})}_{\gamma({\bf q})\tau({\bf t})}\right.\right.\nn \\
&+&\left.\left.\frac{1}{4}
C^{\tilde{\rho}({\bf r})}_{\nu({\bf n})\tilde{\mu}({\bf m})}C^{\tilde{\mu}({\bf m})}_{\gamma({\bf q})
\tau({\bf t})}\right)l^\nu({\bf n})l^\gamma({\bf q})+...\right\}\frac{\partial}{\partial s^\rho({\bf r})}
+\left\{\frac{1}{2}C^\varphi_{\mu({\bf m})\tau({\bf t})}l^\mu({\bf m})\right.
\nn\\
&+&\left.\left(\frac{1}{8}
C^\varphi_{\nu({\bf n})\mu({\bf m})}C^{\mu({\bf m})}_{\gamma({\bf q})\tau({\bf t})}+\frac{1}{4}
C^\varphi_{\nu({\bf n})\tilde{\mu}({\bf m})}C^{\tilde{\mu}({\bf m})}_{\gamma({\bf q})\tau({\bf t})}\right)
l^\nu(n)l^\gamma({\bf q})
+...º\right\}\frac{\partial}{\partial \varphi} \nn \\
\tilde{X}^L_{s^\tau({\bf t})}&=&\frac{\partial}{\partial l^\tau({\bf t})}+
\left\{C^{\tilde{\rho}({\bf r})}_{\mu({\bf m})\tilde{\tau}({\bf t})}l^\mu(m)+
\frac{1}{2}C^{\tilde{\rho}({\bf r})}_{\nu({\bf n})\tilde{\mu}({\bf m})}
C^{\tilde{\mu}({\bf m})}_{\gamma({\bf q})
\tilde{\tau}({\bf t})}l^\nu({\bf n})l^\gamma({\bf q})+...\right\}
\frac{\partial}{\partial s^\rho({\bf r})} \nn \\
&+&\left\{C^\varphi_{\mu({\bf m})\tilde{\tau}({\bf t})}l^\mu({\bf m})+\frac{1}{2}
C^\varphi_{\nu({\bf n})\tilde{\mu}({\bf m})}C^{\tilde{\mu}({\bf m})}
_{\gamma({\bf q})
\tilde{\tau}({\bf t})}l^\nu({\bf n})l^\gamma({\bf q})+...\right\}
\frac{\partial}{\partial \varphi}  \\
\tilde{X}^L_{a({\bf t})}&=&\frac{\partial}{\partial a({\bf t})} 
+\left\{C^{\bf r}_{\mu({\bf m}){\bf t}}l^\mu({\bf m})+
\frac{1}{2}C^{\bf r}_{\nu({\bf n}){\bf m}} C^{\bf m}_{\gamma({\bf q}){\bf t}}
l^\nu({\bf n})l^\gamma({\bf q})+...\right\}\frac{\partial}{\partial a({\bf r})} 
 \nn \\
&+&\left\{ \frac{1}{2} C^{\tilde{\rho}({\bf r})}_{{\bf m}{\bf t}}a({\bf m})+
\frac{1}{4}C^{\tilde{\rho}({\bf r})}_{{\bf q}{\bf m}}
C^{\bf m}_{\nu({\bf n}){\bf t}}l^\nu({\bf n})a({\bf q})+...\right\}\frac{\partial}{\partial s^\rho({\bf p})}\nn \\
&+&\left\{\frac{1}{2}C^\varphi_{{\bf m}{\bf t}}a({\bf m})+C^\varphi_{\mu({\bf m}){\bf t}}l^\mu({\bf m})+
\frac{1}{2}C^\varphi_{\rho({\bf r}){\bf m}}C^{\bf m}_{\nu({\bf n}){\bf t}}
l^\rho({\bf r})l^\nu({\bf n})\nn\right. \\
&+&\left.\frac{1}{4}C^\varphi_{{\bf n}{\bf m}}C^{\bf m}_{\rho({\bf r}){\bf t}}
l^\rho({\bf r})a({\bf n})+...\right\}
\frac{\partial}{\partial \varphi} \nn \\
\tilde{X}^L_\varphi&=& \frac{\partial}{\partial \varphi} \nn \ \ ,
\eea
and also the right-invariant vector fields
$\tilde{X}^R_{(k)}(g)=\frac{\partial g''^l(g',g)}{\partial g'^{(k)}}|_{
g'=e,g=g}\frac{\partial}{\partial g^l}$,
\bea
\tilde{X}^R_{l^\tau({\bf t})}&=&\frac{\partial}{\partial l^\tau({\bf t})}+
\frac{1}{2}C^{\rho({\bf r})}_{\tau({\bf t})\nu({\bf n})} l^\nu({\bf n})
\frac{\partial}{\partial l^\rho ({\bf r})} +
\left(\frac{1}{2}C^{\tilde{\rho}({\bf r})}_{\tau({\bf t})\nu({\bf n})} 
l^\nu({\bf n})+C^{\tilde{\rho}({\bf r})}_{\tau({\bf t})\tilde{\nu}({\bf n})} 
s^\nu({\bf n})\right)\frac{\partial}{\partial s^\rho ({\bf r})}  \nn \\
&+&C^{\bf r}_{\tau({\bf t}){\bf n}}a({\bf n})\frac{\partial}
{\partial a({\bf r})} + \left(\frac{1}{2}C^\varphi_{\tau({\bf t})\nu({\bf n})} 
l^\nu({\bf n})+C^\varphi_{\tau({\bf t})\tilde{\nu}({\bf n})} 
s^\nu({\bf n})+C^\varphi_{\tau({\bf t}){\bf n}}a({\bf n})\right)
\frac{\partial}{\partial \varphi} \nn \\
\tilde{X}^R_{s^\tau({\bf t})}&=&\frac{\partial}{\partial s^\tau({\bf t})} \\
\tilde{X}^R_{a({\bf t})}&=&\frac{\partial}{\partial a({\bf t})}+
\frac{1}{2}C^{\tilde{\rho}({\bf r})}_{\bf tn}a({\bf n})
\frac{\partial}{\partial s^\rho({\bf r})}+\frac{1}{2}C^\varphi_{\bf tn}
a({\bf n})\frac{\partial}{\partial \varphi} \nn \\
\tilde{X}^R_\varphi&=&\frac{\partial}{\partial \varphi} \nn \ \ .
\eea
Analogously, the quantization one-form, $\Theta={\theta^L}^{(\varphi)}=
\frac{\partial \varphi''^l(g',g)}{\partial g^{(k)}}|_{g=g,
g'=g^{-1}}dg^k$ has the expression
\bea
\Theta&=&{\theta^L}^{(\varphi)}
=d\varphi\nn \\
&+&\left\{-\frac{1}{2}
C^\varphi_{\mu({\bf m})\nu({\bf n})}l^\mu({\bf m})+\left(\frac{1}{8}C^\varphi_
{\rho({\bf r})\mu({\bf m})}C^{\mu({\bf m})}_{\gamma({\bf q})\nu({\bf n})}+
\frac{1}{4}C^\varphi_
{\rho({\bf r})\tilde{\mu}({\bf m})}C^{\tilde{\mu}({\bf m})}_{\gamma({\bf q})
\nu({\bf n})}\right)l^\rho({\bf r})l^\gamma({\bf q})\right\}dl^\nu({\bf n}) \nn \\
&+&\left\{-C^\varphi_{\mu({\bf m})\tilde{\nu}({\bf n})}l^\mu({\bf m})+
\frac{1}{2}C^\varphi_
{\rho({\bf r})\tilde{\mu}({\bf m})}C^{\tilde{\mu}({\bf m})}
_{\gamma({\bf q})\tilde{\nu}({\bf n})}l^\rho({\bf r})l^\gamma({\bf q})\right\}
ds^\nu({\bf n}) \\
&+&\left\{-\frac{1}{2}C^\varphi_{\bf mn}a({\bf m})
-C^\varphi_{\mu({\bf m}){\bf n}}
l^\mu({\bf m})
+\frac{1}{4}\left(2C^\varphi_{\bf mn}C^{\bf m}_{\rho({\bf r}){\bf q}}
+C^\varphi_{\bf qm}C^{\bf m}_{\rho({\bf r}){\bf n}}\right)l^\rho({\bf r})
a({\bf q}) \right. \nn \\
&+&\left.\frac{1}{2}C^\varphi_{\rho({\bf r}){\bf m}}C^{\bf m}_{\gamma({\bf q})
{\bf n}}
l^\rho({\bf r})l^\gamma({\bf q})\right\} da({\bf n}) \nn \ \ .
\eea

\subsection{Semi-classical formalism}
Except for critical combinations of the pseudo-cohomology parameters
(related to possible anomalies) which 
we are not going to focus on here, the simultaneous kernel of $\Theta$ and
$d\Theta$ is spanned by 
$\langle\tilde{X}^L_{l^\mu({\bf 0})}, \tilde{X}^L_{s^\mu({\bf 0})}, 
\tilde{X}^L_{a({\bf 0})}\rangle$, and denoted by 
${\cal G}_\Theta$. These vector fields are the generators of the generalized
equations of motion, the classical phase space being  given
by $G/G_\Theta$. The Noether invariants under these classical 
trajectories are
\bea
{\cal N}_{l^\tau({\bf t})}&=& i_{\tilde{X}^R_{l^\tau({\bf t})}}\Theta \nn \\
&=& C^\varphi_{\tau({\bf t})\mu({\bf m})}l^\mu({\bf m})+
C^\varphi_{\tau({\bf t})\tilde{\mu}({\bf m})}s^\mu({\bf m})+
C^\varphi_{\tau({\bf t}){\bf m}}a({\bf m}) \nn \\
&+&\frac{3}{4}\left(\frac{1}{2}C^\varphi_{\rho({\bf r})\mu({\bf m})}
C^{\mu({\bf m})}_{\sigma({\bf s})\tau({\bf t})}+
C^\varphi_{\rho({\bf r})\tilde{\mu}({\bf m})}C^{\tilde{\mu}({\bf m})}_
{\gamma({\bf q})\tau({\bf t})}\right)l^\rho({\bf r})l^\gamma({\bf q})\nn \\
&+&C^\varphi_{\rho({\bf r})\tilde{\mu}({\bf m})}C^{\tilde{\mu}({\bf m})}_
{\tilde{\gamma}({\bf q})\tau({\bf t})} l^\rho({\bf r})s^\gamma({\bf q})+
\frac{1}{2}C^\varphi_{\bf rm}C^{\bf m}_{{\bf q}\tau({\bf t})}a({\bf r})
a({\bf q})+ C^\varphi_{\rho({\bf r}){\bf m}}C^{\bf m}_{{\bf q}\tau({\bf t})}
l^\rho({\bf r})a({\bf q})+...\nn \\ 
{\cal N}_{s^\tau({\bf t})}&=&i_{\tilde{X}^R_{s^\tau({\bf t})}}\Theta \nn \\
&=&C^\varphi_{\tilde{\tau}({\bf t})\mu({\bf m})} l^\mu({\bf m})
+\frac{1}{2}C^\varphi_{\rho({\bf r})\tilde{\mu}({\bf m})}
C^{\tilde{\mu}({\bf m})}_{\gamma({\bf q})\tilde{\tau}({\bf t})}l^\rho({\bf r})
l^\gamma({\bf q})+... \nn \\
{\cal N}_{a({\bf t})}&=&i_{\tilde{X}^R_{a({\bf t})}}\Theta  \\
&=& C^\varphi_{\bf tm}a({\bf m})+C^\varphi_{{\bf t}\mu({\bf m})}
l^\mu({\bf m}) \nn \\
&+&\frac{1}{4}\left(2C^\varphi_{\bf mt}C^{\bf m}_{\rho({\bf r}){\bf q}}+
C^\varphi_{\bf qm}C^{\bf m}_{\rho({\bf r}){\bf t}}+
2C^\varphi_{\rho({\bf r})\tilde{\mu}({\bf m})}
C^{\tilde{\mu}({\bf m})}_{\bf qt}\right)l^\rho({\bf r})a({\bf q}) \nn \\
&+&\frac{1}{2}C^\varphi_{\rho({\bf r}){\bf m}}
C^{\bf m}_{\gamma({\bf q}){\bf t}}l^\rho({\bf r})l^\gamma({\bf q})+... \nn 
\ \ .
\eea
Finally, the Noether invariants associated with the vector 
fields inside ${\cal G}_\Theta$ can be written in terms of the remaining 
(basic) Noether invariants, which can be used as coordinates in the solution 
manifold. Up to the second order (inherited from the third order in the
group law) we find,
\bea
{\cal N}_{l^\tau({\bf 0})}&=&\left\{\frac{3}{4}
\left(\frac{1}{2}
C^\varphi_{\rho({\bf r})\mu({\bf m})}C^{\mu({\bf m})}_{\gamma({\bf q})
\tau({\bf 0})}+C^\varphi_{\rho({\bf r})\tilde{\mu}({\bf m})}
C^{\tilde{\mu}({\bf m})}_{\gamma({\bf q})\tau({\bf 0})}\right)
(C^\varphi_{\tilde{\nu}({\bf n})\rho({\bf r})})^{-1}
(C^\varphi_{\tilde{\nu}'({\bf n}')\gamma({\bf q})})^{-1}\right. \nn \\
&-&C^\varphi_{\rho({\bf r})\tilde{\mu}({\bf m})}
C^{\tilde{\mu}({\bf m})}_{\tilde{\gamma}({\bf q})\tau({\bf 0})}
(C^\varphi_{\tilde{\nu}({\bf n})\rho({\bf r})})^{-1}
(C^\varphi_{\alpha({\bf a})\tilde{\gamma}({\bf q})})^{-1}
\left(C^\varphi_{\alpha({\bf a})\gamma({\bf q})}
(C^\varphi_{\tilde{\nu}'({\bf n}')\gamma'({\bf q}')})^{-1} \right.\nn \\
&-&\left.C^\varphi_{\alpha({\bf a}){\bf q}'}(C^\varphi_{\bf a'q'})^{-1}
C^\varphi_{{\bf a}'\rho'({\bf r}')}(C^\varphi_{\tilde{\nu}'({\bf n}')
\rho'({\bf r}')})^{-1}\right) \nn \\
&+&\frac{1}{2}C^\varphi_{\bf rm}C^{\bf m}_{{\bf s}\tau({\bf 0})}
(C^\varphi_{\bf t' r})^{-1}C^\varphi_{{\bf t}'\gamma'({\bf q}')}
(C^\varphi_{\tilde{\nu}({\bf n})\gamma'({\bf q'})})^{-1}
(C^\varphi_{\bf t''s})^{-1}C^\varphi_{{\bf t''}\gamma''({\bf q}'')}
(C^\varphi_{\tilde{\nu'}({\bf n'})\gamma''({\bf q''})})^{-1}\nn \\
&-&\left.C^\varphi_{\rho({\bf r})\mu}C^\mu_{{\bf s}\tau({\bf 0})}
(C^\varphi_{\tilde{\nu}({\bf n})\rho({\bf r})})^{-1}
(C^\varphi_{\bf t''s})^{-1}C^\varphi_{{\bf t''}\gamma''({\bf q''})}
(C^\varphi_{\tilde{\nu}'({\bf n}')\gamma''({\bf q''})})^{-1}\right\}
{\cal N}_{s^\nu({\bf n})}{\cal N}_{s^{\nu'}({\bf n'})} \nn \\
&+&C^\varphi_{\rho({\bf r})\tilde{\mu}({\bf m})}
C^{\tilde{\mu}({\bf m})}_{\tilde{\gamma}({\bf q})\tau({\bf 0})}
(C^\varphi_{\tilde{\nu}({\bf n})\rho({\bf r})})^{-1}
(C^\varphi_{\alpha({\bf a})\tilde{\gamma}({\bf q})})^{-1}
{\cal N}_{s^\nu({\bf n})}{\cal N}_{l^\alpha({\bf a})}\nn \\
&+&\left\{-C^\varphi_{\rho({\bf r})\tilde{\mu}({\bf m})}
C^{\tilde{\mu}({\bf m})}_{\tilde{\gamma}({\bf q})\tau({\bf 0})}
(C^\varphi_{\tilde{\nu}({\bf n})\rho({\bf r})})^{-1}
(C^\varphi_{\alpha({\bf a})\tilde{\gamma}({\bf q})})^{-1}
C^\varphi_{\alpha({\bf a}){\bf q}'}(C^\varphi_{\bf n'q'})^{-1}\right.\nn \\
&-&\frac{1}{2}C^\varphi_{\bf rm}C^{\bf m}_{{\bf s}\tau({\bf 0})}
(C^\varphi_{\bf n' r})^{-1}(C^\varphi_{\bf t''s})^{-1}
C^\varphi_{{\bf t}''\gamma''({\bf q}'')}
(C^\varphi_{\tilde{\nu}({\bf n})\gamma''({\bf q}'')})^{-1} \label{Noether} \\
&+&\left.(C^\varphi_{\bf n's})^{-1}C^\varphi_{{\bf s}\gamma'({\bf q}')}
(C^\varphi_{\tilde{\nu}({\bf n})\gamma'({\bf q}')})^{-1}+
C^\varphi_{\rho({\bf r})\mu}C^\mu_{{\bf s}\tau({\bf 0})}
(C^\varphi_{\tilde{\nu}({\bf n})\rho({\bf r})})^{-1}
(C^\varphi_{\bf n's})^{-1}\right\}\nn\\
&\cdot& {\cal N}_{s^\nu({\bf n})}{\cal N}_{a({\bf n}')}
+\frac{1}{2}C^\varphi_{\bf rm}C^{\bf m}_{{\bf q}\tau({\bf 0})}
(C^\varphi_{\bf t'q})^{-1}(C^\varphi_{\bf t''r})^{-1}
{\cal N}_{a({\bf t}')}{\cal N}_{a({\bf t}'')}+...\nn \\
{\cal N}_{s^\tau({\bf 0})}&=&
\frac{1}{2}C^\varphi_{\rho({\bf r})\tilde{\mu}({\bf m})}
C^{\tilde{\mu}({\bf m})}_{\gamma({\bf q})\tilde{\tau}({\bf 0})}
(C^\varphi_{\tilde{\tau}'({\bf t}')\rho({\bf r})})^{-1}
(C^\varphi_{\tilde{\tau}''({\bf t}'')\gamma({\bf q})})^{-1}
{\cal N}_{s^{\tau'}({\bf t'})}{\cal N}_{s^{\tau''}({\bf t''})}+... \nn\\
{\cal N}_{a({\bf 0})}&=&
\left\{-\frac{1}{4}\left(2C^\varphi_{\bf m0}C^{\bf m}_{\rho({\bf r}){\bf q}}+
C^\varphi_{\bf qm}C^{\bf m}_{\rho({\bf r}){\bf 0}}
+2C^\varphi_{\rho({\bf r})\tilde{\mu}({\bf m})}
C^{\tilde{\mu}({\bf m})}_{\bf q0}\right)\right.\nn \\
&&(C^\varphi_{\tilde{\tau}'({\bf t}')\rho({\bf r})})^{-1}
(C^\varphi_{\bf t''q})^{-1}C^\varphi_{{\bf t}''\gamma''({\bf q}'')}
(C^\varphi_{\tilde{\nu}''({\bf n}'')\gamma''({\bf q}'')})^{-1} \nn \\
&+&\left.\frac{1}{2} C^\varphi_{\rho({\bf r}){\bf m}}
C^{\bf m}_{\gamma({\bf q}){\bf 0}}
(C^\varphi_{\tilde{\tau}'({\bf t}')\rho({\bf r})})^{-1}
(C^\varphi_{\tilde{\nu}''({\bf n}'')\gamma({\bf q})})^{-1}\right\}
{\cal N}_{s^{\tau'}({\bf t}')}{\cal N}_{s^{\nu''}({\bf n}'')}\nn \\
&+&\left\{\frac{1}{4}\left(2C^\varphi_{\bf m0}C^{\bf m}_{\rho({\bf r}){\bf q}}+
C^\varphi_{\bf qm}C^{\bf m}_{\rho({\bf r}){\bf 0}}+
2C^\varphi_{\rho({\bf r})\tilde{\mu}({\bf m})}
C^{\tilde{\mu}({\bf m})}_{\bf q0}\right)
(C^\varphi_{\tilde{\tau}'({\bf t}')\rho({\bf r})})^{-1}
(C^\varphi_{\bf t''q})^{-1}\right\}\nn\\
&\cdot&{\cal N}_{s^{\tau'}({\bf t}')}
{\cal N}_{a({\bf t}'')}+... \nn 
\eea

\subsection{Quantum analysis}
The regular representation is originally realized on those complex functions 
defined on the extended group which satisfy the $U(1)-$equivariance 
condition
\bea
\tilde{X}^R_\varphi \Psi=\frac{\partial}{\partial\varphi}\Psi=i\Psi \ \ ,
\eea 
and is infinitesimally implemented by the right-invariant
vector fields. In order to reduce this representation, we must impose 
a polarization condition associated with a (left-invariant)  
polarization algebra ${\cal G}_{\cal P}$, 
implying the selection of a particular class of representations.
 
We shall implement the easiest polarization one can conceive in this context.
In order to do so, we introduce the following notation: given a vector 
${\bf m}\neq 0$, let us call $m_f$ the first non-zero entry 
$m_i$. For ${\bf m}=0$, let $m_f=0$. 
Now we introduce the following polarization:
\bea
{\cal G}_\Theta= \langle{X^L_{l^\mu({\bf t})}}_{(t_f\leq 0)},
{X^L_{s^\mu({\bf t})}}_{(t_f\leq 0)}, {X^L_{a({\bf t})}}_{(t_f\leq 0)}\rangle 
\ \ ,
\eea
defining the polarization equations:
\bea
{X^L_{l^\mu({\bf t})}}_{(t_f\leq 0)}\Psi=0 \nn \\
{X^L_{s^\mu({\bf t})}}_{(t_f\leq 0)}\Psi=0  \\
{X^L_{a({\bf t})}}_{(t_f\leq 0)}\Psi=0 \nn \ \ .
\eea
A general solution can be written as a linear combination:
\bea
\Psi&=&A|0\rangle+\sum_{{\bf m}(m_f>0)} \left(A_{\mu({\bf m})}|l^\mu({\bf m})\rangle+
A_{\tilde{\mu}({\bf m})}|s^\mu({\bf m})\rangle+
A_{\bf m}|a({\bf m})\rangle\right) \nn \\
&+&\sum \left(A_{\mu({\bf m})\nu({\bf n})}|l^\mu({\bf m})l^\nu({\bf n})\rangle+
A_{\mu({\bf m})\tilde{\nu}({\bf n})}|l^\mu({\bf m})s^\nu({\bf n})\rangle\right.  \\
&+&A_{\mu({\bf m}){\bf n}}|l^\mu({\bf m})a({\bf n})\rangle+
A_{\tilde{\mu}({\bf m})\tilde{\nu}({\bf n})}|s^\mu({\bf m})s^\nu({\bf n})
\rangle \nn \\
&+&\left.A_{\tilde{\mu}({\bf m}){\bf n}}|s^\mu({\bf m})a({\bf n})\rangle+
A_{{\bf m}{\bf n}}|a({\bf m})a({\bf n})\rangle\right) \nn \ \ ,
\eea
where the functions in the basis, $|...\rangle$, are already polarized and 
present in fact a product structure 
in terms of the functions $\{...\}$ that we are giving below; 
they share a common weight 
function $W$ depending on the representation,
\bea
|0\rangle&=&\zeta W \nn\\
|l^\mu({\bf m})\rangle=\zeta W \{l^\mu({\bf m})\},\ \ \
|s^\mu({\bf m})\rangle&=&\zeta W \{s^\mu({\bf m})\}, \ \ \
|a({\bf m})\rangle=\zeta W \{a({\bf m})\}\\
|l^{\mu_1}({\bf m}_i)...
s^{\mu_1}({\bf m}_j)...
a({\bf m}_k)...\rangle&=&\zeta W
\{l^{\mu_1}({\bf m}_i)\}...
\{s^{\mu_1}({\bf m}_j)\}...
\{a({\bf m}_k)\}... \nn \ \ .
\eea
The explicit forms of the corresponding functions are,
\bea
W&=&1-\sum_{{\bf n}_{(n_f>0)}} \left(\frac{1}{2}C^\varphi_{\nu({\bf n})\mu({\bf m})}
l^\mu({\bf m})l^\nu({\bf n})+\frac{1}{2}C^\varphi_{\bf mn}a({\bf m})a({\bf n})
\nn\right. \\
&+&\left.C^\varphi_{\mu({\bf m})\tilde{\nu}({\bf n})}l^\mu({\bf m})
s^\nu({\bf n})
+C^\varphi_{\mu({\bf m}){\bf n}}l^\mu({\bf m})a({\bf n})+...\right) \nn \\
\{l^\gamma({\bf q})\}&=&l^\gamma({\bf q})-\sum_{t_f\leq0, m_f>0,n_f>0,m_f+
n_f\geq q_f}\left[\frac{1}{2}C^{\gamma({\bf q})}_{\mu({\bf m})\tau({\bf t})}
l^\mu({\bf m})l^\tau({\bf t})\right.\nn\\
&+&\frac{1}{8}\left(-\sum_{r_f\leq0}
C^{\rho({\bf r})}_{\mu({\bf m})\tau({\bf t})}
C^{\gamma({\bf q})}_{\nu({\bf n})\rho({\bf r})}
+\sum_{r_f>0}\left(C^{\gamma({\bf q})}_{\rho({\bf r})\tau({\bf t})}
C^{\rho({\bf r})}_{\mu({\bf m})\nu({\bf n})}\right.\right.\nn\\
&+&\left.\left.\left.
C^{\gamma({\bf q})}_{\mu({\bf m})\rho({\bf r})}
C^{\rho({\bf r})}_{\nu({\bf n})\tau({\bf t})}\right)\right)l^\tau({\bf t})
l^\mu({\bf m})l^\nu({\bf n})+...\right] \nn \\
\{s^\gamma({\bf q})\}&=&s^\gamma({\bf q})-\sum_{t_f\leq0, m_f>0,n_f>0,m_f+
n_f\geq q_f}\left(\frac{1}{2}C^{\tilde{\gamma}({\bf q})}_{\mu({\bf m})\tau({\bf t})}
l^\mu({\bf m})l^\tau({\bf t})+
C^{\tilde{\gamma}({\bf q})}_{\mu({\bf m})\tilde{\tau}({\bf t})}
l^\mu({\bf m})s^\tau({\bf t})\right. \nn \\
&+&\left.\frac{1}{2}C^{\tilde{\gamma}({\bf q})}_{\bf mt})
a({\bf m})a({\bf t})\right)
+\left[\sum_{r_f\leq0}\left(\frac{1}{8}C^{\rho({\bf r})}_{\mu({\bf m})\tau({\bf t})}
C^{\tilde{\gamma}({\bf q})}_{\nu({\bf n})\rho({\bf r})}-
\frac{1}{2}C^{\tilde{\rho}({\bf r})}_{\mu({\bf m})\tau({\bf t})}
C^{\tilde{\gamma}({\bf q})}_{\nu({\bf n})\tilde{\rho}({\bf r})}\right) 
\right. \\
&+&\left.\sum_{r_f>0}\left(-\frac{1}{8}C^{\tilde{\gamma}({\bf q})}
_{\rho({\bf r})\tau({\bf t})}
C^{\rho({\bf r})}_{\mu({\bf m})\nu({\bf n})}+\frac{1}{4}
C^{\tilde{\gamma}({\bf q})}_{\mu({\bf m})\rho({\bf r})}
C^{\rho({\bf r})}_{\nu({\bf n})\tau({\bf t})}\right)\right]
l^\tau({\bf t})l^\mu({\bf m})
l^\nu({\bf n}) \nn \\
&+&\frac{1}{2}\left[\sum_{r_f\leq0}C^{\tilde{\rho}({\bf r})}_{\mu({\bf m})
\tilde{\tau}({\bf t})}C^{\tilde{\gamma}({\bf q})}_{\nu({\bf n})
\tilde{\rho}({\bf r})}+\sum_{r_f>0}\left(\frac{1}{2}
C^{\tilde{\gamma}({\bf q})}_{\rho({\bf r})
\tilde{\tau}({\bf t})}C^{\rho({\bf r})}_{\mu({\bf m})
{\nu}({\bf n})}\right.\right. \nn \\
&+&\left.\left.C^{\tilde{\gamma}({\bf q})}_{\mu({\bf m})
\tilde{\rho}({\bf r})}C^{\tilde{\rho}({\bf r})}_{\nu({\bf n})
\tilde{\tau}({\bf t})}\right)\right]l^\mu({\bf m})l^\nu({\bf n})s^\tau({\bf t})+\nn \\
&-&\frac{1}{4}\left[\sum_{r_f\leq0}\left(C^{\bf r}_{\mu({\bf m}){\bf t}}
C^{\tilde{\gamma}({\bf q})}_{\bf nr}+2C^{\tilde{\rho}({\bf r})}_{\bf nt}
C^{\tilde{\gamma}({\bf q})}_{\mu({\bf m})\tilde{\rho}({\bf r})}\right)
-\sum_{r_f>0}C^{\tilde{\gamma}({\bf q})}_{\bf nr}
C^{\bf r}_{\mu({\bf m}){\bf t}}\right]
l^\mu({\bf m})a({\bf t})a({\bf n})+... \nn \\
\{a({\bf q})\}&=& a({\bf q})-\sum_{t_f\leq0, m_f>0,n_f>0,m_f+
n_f\geq q_f}(C^{\bf q}_{\mu({\bf m}){\bf t}}l^{\mu({\bf m})}a({\bf t})
-\frac{1}{2}\left[\sum_{r_f\leq0}C^{\bf r}_{\mu({\bf m}){\bf t}}
C^{\bf q}_{\nu({\bf n}){\bf r}} \right.\nn \\
&+&\left.\sum_{r_f>0}\left(\frac{1}{2}C^{\bf q}_{\rho({\bf r}){\bf t}}C^{\rho({\bf r})}_
{\mu({\bf m})\nu({\bf n})}-C^{\bf q}_{\mu({\bf m}){\bf r}}C^{\bf r}
_{\nu({\bf n}){\bf t}}\right)\right]l^\mu({\bf m})l^{\nu({\bf n})}a({\bf t})+...) \nn \ \ .
\eea
The action of the starting algebra is implemented by the right-invariant
vector fields in such a way that those vector fields with $m_f\leq 0$ act
as raising operators and those with $m_f>0$ are lowering operators.
Finally, the explicit action of these right-invariant vector
fields on the lowest-order functions of the reduced Hilbert space is:
\bea
\tilde{X}^R_{a({\bf t})_{(t_f\leq0)}}|0\rangle&=&
C^\varphi_{\bf tn}|a({\bf n})\rangle+C^\varphi_{{\bf t}\nu({\bf n})}
|l^{\nu({\bf n})}\rangle-\frac{1}{2}\sum_{r_f\leq0}C^{\tilde{\rho}({\bf r})}_{\bf tn}
C^\varphi_{\mu({\bf m})\tilde{\rho}({\bf r})}|a({\bf n})l^\mu({\bf m})\rangle+... 
\nn \\
\tilde{X}^R_{a({\bf t})_{(t_f\leq0)}}|l^\gamma({\bf q})\rangle&=&
C^\varphi_{\bf tn}|a({\bf n})l^\gamma({\bf q})\rangle+
C^\varphi_{{\bf t}\nu({\bf n})}|l^\nu({\bf n})l^\gamma({\bf q})\rangle+... \nn \\
\tilde{X}^R_{a({\bf t})_{(t_f\leq0)}}|s^\gamma({\bf q})\rangle&=&
C^{\tilde{\gamma}({\bf q})}_{\bf tn}|a({\bf n})\rangle
+C^\varphi_{\bf tn}|a({\bf n})s^\gamma({\bf q})\rangle+
C^\varphi_{{\bf t}\nu({\bf n})}
|l^\nu({\bf n})s^\gamma({\bf q})\rangle\nn \\
&+&\left[\sum_{r_f\leq0}(\frac{1}{4}C^{\bf r}_{\mu({\bf m}){\bf t}}
C^{\tilde{\gamma}({\bf q})}_{\bf nr}+
C^{\tilde{\rho}({\bf r})}_{\bf nt}C^{\tilde{\gamma}({\bf q})}_
{\mu({\bf m})\tilde{\rho}({\bf r})})
\nn \right.\\
&-&\left.\sum_{r_f>0}\frac{1}{4}
C^{\bf r}_{\mu({\bf m}){\bf t}}C^{\tilde{\gamma}({\bf q})}_{\bf nr})\right]
|l^\mu({\bf m})a({\bf n})\rangle+... \nn \\
\tilde{X}^R_{a({\bf t})_{(t_f\leq0)}}|a({\bf q})\rangle&=&
C^\varphi_{\bf tn}|a({\bf n})a({\bf q})\rangle+ C^\varphi_{{\bf t}\nu({\bf n})}
|l^\nu({\bf n})a({\bf q})\rangle+\frac{1}{2}C^{\bf q}_{{\bf t}\mu({\bf m})}
|l^\mu({\bf m})\rangle\nn \\
&-&\frac{1}{2}\left[\sum_{r_f\leq0}C^{\bf r}_{{\bf t}\mu({\bf m})}
C^{\bf q}_{\nu({\bf n}){\bf r}}+
\sum_{r_f>0}C^{\bf q}_{\mu({\bf m}){\bf r}}C^{\bf r}_{\nu({\bf n}){\bf t}}
\nn \right.\\
&-&\left.\frac{1}{2}C^{\bf q}_{\rho({\bf r}){\bf t}}C^{\rho({\bf r})}
_{\mu({\bf m})\nu({\bf n})}\right]
|l^\mu({\bf m})l^\nu({\bf n})\rangle+... \nn \\
\tilde{X}^R_{s^\tau({\bf t})_{(t_f\leq0)}}|0\rangle&=&
C^\varphi_{\tilde{\tau}({\bf t})\nu({\bf n})}|l^\nu({\bf n})\rangle+... \nn \\
\tilde{X}^R_{s^\tau({\bf t})_{(t_f\leq0)}}|l^\gamma({\bf q})\rangle&=&
C\varphi_{\tilde{\tau}({\bf t})\nu({\bf n})}|l^\nu({\bf n})
l^\gamma({\bf q})\rangle+... \nn \\
\tilde{X}^R_{s^\tau({\bf t})_{(t_f\leq0)}}|s^\gamma({\bf q})\rangle&=&
C^{\tilde{\gamma}({\bf q})}_{\tilde{\tau}({\bf t})\mu({\bf m})}
|l^\mu({\bf m})\rangle+C^\varphi
_{\tilde{\tau}({\bf t})\nu({\bf n})}|l^\nu({\bf n})s^\gamma({\bf q})\rangle \\
&+&\frac{1}{2}\left[\sum_{r_f\leq0}
C^{\tilde{\rho}({\bf r})}_{\mu({\bf m})\tilde{\tau}({\bf t})}
C^{\tilde{\gamma}({\bf q})}_{\nu({\bf n})\tilde{\rho}({\bf r})}+
\sum_{r_f>0}\left(\frac{1}{2}C^{\tilde{\gamma}({\bf q})}
_{\rho({\bf r})\tilde{\tau}({\bf t})}
C^{\rho({\bf r})}_{\mu({\bf m})\nu({\bf n})}\nn \right.\right.\\
&-&\left.\left.C^{\tilde{\gamma}({\bf q})}_{\mu({\bf m})\tilde{\rho}({\bf r})}
C^{\tilde{\rho}({\bf r})}_{\nu({\bf n})\tilde{\tau}({\bf t})}\right)\right]
|l^\mu({\bf m})l^\nu({\bf n})\rangle+... \nn \\
\tilde{X}^R_{s^\tau({\bf t})_{(t_f\leq0)}}|a({\bf q})\rangle&=&
C^\varphi_{\tilde{\tau}({\bf t})\nu({\bf n})}|l^\nu({\bf n})a({\bf q})\rangle+... 
\nn \\
\tilde{X}^R_{l^\tau({\bf t})_{(t_f\leq0)}}|0\rangle&=&
C^\varphi_{\tau({\bf t})\nu({\bf n})}|l^\nu({\bf n})\rangle+
C^\varphi_{\tau({\bf t})\tilde{\nu}({\bf n})}|s^\nu({\bf n})\rangle+
C^\varphi_{\tau({\bf t}){\bf n}}|a({\bf n})\rangle\nn \\
&-&\sum_{r_f\leq0}\left(\frac{1}{4}C^{\rho({\bf r})}_{\tau({\bf t})\nu({\bf n})}
C^\varphi_{\mu({\bf m})\rho({\bf r})}
+\frac{1}{2}C^{\tilde{\rho}({\bf r})}_{\tau({\bf t})\nu({\bf n})}
C^\varphi_{\mu({\bf m})\tilde{\rho}({\bf r})}\right)
|l^\nu({\bf n})l^\mu({\bf m})\rangle \nn \\
&-&\sum_{r_f\leq0}C^{\tilde{\rho}({\bf r})}_{\tau({\bf t})\tilde{\nu}({\bf n})}
C^\varphi_{\mu({\bf m})\tilde{\rho}({\bf r})}|s^\nu({\bf n})l^\mu({\bf m})\rangle-
\frac{1}{2}\sum_{r_f\leq0}C^{\bf r}_{\tau({\bf t}){\bf n}}
C^\varphi_{\bf mr}|a({\bf n})a({\bf m})\rangle\nn\\
&-&\sum_{r_f\leq0}C^{\bf r}_{\tau({\bf t}){\bf n}}C^\varphi_{\mu({\bf m})
{\bf r}}|a({\bf n})l^\mu({\bf m})\rangle+... \nn \\
\tilde{X}^R_{l^\tau({\bf t})_{(t_f\leq0)}}|l^\gamma({\bf q})\rangle&=&
C^{\gamma({\bf q})}_{\tau({\bf t})\mu({\bf m})}|l^\mu({\bf m})\rangle+
C^\varphi_{\tau({\bf t})\nu({\bf n})}|l^\nu({\bf n})l^\gamma({\bf q})\rangle+
C^\varphi_{\tau({\bf t})\tilde{\nu}({\bf n})}|s^\nu({\bf n})l^\gamma({\bf q})\rangle
\nn\\
&+&C^\varphi_{\tau({\bf t}){\bf n}}|a({\bf n})l^\gamma({\bf q})\rangle
-\frac{1}{8}\left[3\sum_{r_f\leq0}C^{\rho({\bf r})}_{\tau({\bf t})\nu({\bf n}})
C^{\gamma({\bf q})}_{\mu({\bf m})\rho({\bf r})}\nn\right.\\
&+&\left.
\sum_{r_f>0}\left(C^{\gamma({\bf q})}_{\mu({\bf m})\rho({\bf r})}
C^{\rho({\bf r})}_{\nu({\bf n})\tau({\bf t})}
-C^{\gamma({\bf q})}_{\rho({\bf r})\tau({\bf t})}C^{\rho({\bf r})}
_{\mu({\bf m})\nu({\bf n})}\right)\right]|l^\mu({\bf m})l^\nu({\bf n})\rangle+... \nn \\
\tilde{X}^R_{l^\tau({\bf t})_{(t_f\leq0)}}|s^\gamma({\bf q})\rangle&=&
C^{\tilde{\gamma}({\bf q})}_{\tau({\bf t})\mu({\bf m})}|l^\mu({\bf m})\rangle
+C^{\tilde{\gamma}({\bf q})}_{\tau({\bf t})\tilde{\nu}({\bf n})}
|s^\nu({\bf n})\rangle \nn \\
&+&C^\varphi_{\tau({\bf t})\nu({\bf n})}|l^\nu({\bf n})s^\gamma({\bf q})\rangle+
C^\varphi_{\tau({\bf t})\tilde{\nu}({\bf n})}|s^\nu({\bf n})
s^\gamma({\bf q})\rangle\nn \\
&-&\left[\sum_{r_f\leq0}\left(\frac{3}{8}
C^{\rho({\bf r})}_{\tau({\bf t})\mu({\bf m})}C^{\tilde{\gamma}({\bf q})}_
{\nu({\bf n})\rho({\bf r})}+
\frac{3}{4}C^{\tilde{\rho}({\bf r})}_{\tau({\bf t})\mu({\bf m})}
C^{\tilde{\gamma}({\bf q})}_{\nu({\bf n})\tilde{\rho}({\bf r})}\right)
\right.\nn\\
&+&\sum_{r_f>0}\left(\frac{1}{8}C^{\tilde{\gamma}({\bf q})}_{\mu({\bf m})
\rho({\bf r})}
C^{\rho({\bf r})}_{\nu({\bf n})\tau({\bf t})} 
+\frac{1}{2}
C^{\tilde{\gamma}({\bf q})}_{\mu({\bf m})\tilde{\rho}({\bf r})}
C^{\tilde{\rho}({\bf r})}_{\nu({\bf n})\tau({\bf t})}\nn\right.\\
&-&\left.\left.
\frac{1}{8}C^{\tilde{\gamma}({\bf q})}_{\rho({\bf r})\tau({\bf t})}
C^{\rho({\bf r})}_{\mu({\bf m})\nu({\bf n})}\right)\right]
|l^\mu({\bf m})l^\nu({\bf n})\rangle + 
C^\varphi_{\tau({\bf t}){\bf n}}|a({\bf n})
s^\gamma({\bf q})\rangle       \nn \\
&+&\sum_{r_f\leq0}\left(C^{\tilde{\rho}({\bf r})}_{\tilde{\nu}({\bf n})
\tau({\bf t})}
C^{\tilde{\gamma}({\bf q})}_{\mu({\bf m})\tilde{\rho}({\bf r})}
|l^\mu({\bf m})s^\nu({\bf n})\rangle   
+\frac{1}{2}C^{\bf r}_{{\bf n}\tau({\bf t})}
C^{\tilde{\gamma}({\bf q})}_{\bf mr}|a({\bf m})a({\bf n})\rangle\right)+... \nn \\
\tilde{X}^R_{l^\tau({\bf t})_{(t_f\leq0)}}|a({\bf q})\rangle&=&
C^{\bf q}_{\tau({\bf t}){\bf n}}|a({\bf n})\rangle+ 
C^\varphi_{\tau({\bf t})\nu({\bf n})}|l^\nu({\bf n})a({\bf q})\rangle
+\frac{1}{2}C^\varphi_{\tau({\bf t})\tilde{\nu}({\bf n})}|s^\nu({\bf n})
a({\bf q})\rangle \nn \\
&+& C^\varphi_{\tau({\bf t}){\bf n}}|a({\bf n})a({\bf q})\rangle-
\sum_{r_f\leq0}C^{\bf r}_{\tau({\bf t}){\bf n}}
C^{\bf q}_{\mu({\bf m}){\bf r}}|a({\bf n})l^\mu({\bf m})\rangle+... \nn \\
\tilde{X}^R_{a({\bf t})_{(t_f>0)}}|0\rangle&=&0 \nn \\
\tilde{X}^R_{a({\bf t})_{(t_f>0)}}|a({\bf q})\rangle&=&\delta({\bf q}-{\bf t})|0\rangle
\nn \\
\tilde{X}^R_{a({\bf t})_{(t_f>0)}}|l^\gamma({\bf q})\rangle&=&0 \nn \\
\tilde{X}^R_{a({\bf t})_{(t_f>0)}}|s^\gamma({\bf q})\rangle&=&0 \nn \\
\tilde{X}^R_{s^\tau({\bf t})_{(t_f>0)}}|0\rangle&=&0 \nn \\
\tilde{X}^R_{s^\tau({\bf t})_{(t_f>0)}}|a({\bf q})\rangle&=& 0 \nn \\
\tilde{X}^R_{s^\tau({\bf t})_{(t_f>0)}}|l^\gamma({\bf q})\rangle&=&0 \nn \\
\tilde{X}^R_{s^\tau({\bf t})_{(t_f>0)}}|s^\gamma({\bf q})\rangle&=&
\delta^\gamma_\tau\delta({\bf q}-{\bf t})|0\rangle \nn \\
\tilde{X}^R_{l^\tau({\bf t})_{(t_f>0)}}|0\rangle&=&0 \nn \\
\tilde{X}^R_{l^\tau({\bf t})_{(t_f>0)}}|a({\bf q})\rangle&=& 0 \nn \\
\tilde{X}^R_{l^\tau({\bf t})_{(t_f>0)}}|l^\gamma({\bf q})\rangle&=&\delta^\gamma_
\tau\delta({\bf q}-{\bf t})|0\rangle\nn \\
\tilde{X}^R_{l^\tau({\bf t})_{(t_f>0)}}|s^\gamma({\bf q})\rangle&=&0 \nn \ \ , 
\eea
where $|0\rangle$ behaves  as a vacuum state and is unique by 
construction.
\ni The action on higher-order functions is limited by the initial perturbative
development in the group law. Apart from that, its computation is 
completely straightforward, though highly tedious. 

As we said in the {\bf Introduction}, the Hilbert product 
$\langle ...|...\rangle$
is introduced by imposing $\langle 0|0\rangle=1$ and specifying the adjoint 
for each operator. In order to do that, we set the notation
\bea
\hat{L}_\mu({\bf m})=\tilde{X}^R_{l^\mu({\bf m})} , \
\hat{A}({\bf m})=\tilde{X}^R_{a({\bf m})} \ , \
\hat{S}^\rho({\bf m})=\tilde{X}^R_{s_\rho({\bf m})}\nn \ \ , 
\eea
and stablish the corresponding adjoint operators
\bea
\hat{L}^\dagger_\mu({\bf m})=\hat{L}_\mu(-{\bf m}) \ , \ 
\hat{A}^\dagger({\bf m})= \hat{A}(-{\bf m}) \ , \
(\hat{S}^\rho)^\dagger({\bf m})=  \hat{S}^\rho(-{\bf m})  \nn \ \ .
\eea

\section{Physical interpretation}

As we mentioned in the {\bf Introduction}, in the higher-dimensional 
case we no longer have
the connection with Physics provided by the Polyakov action. This is the
reason why we have not developed the analysis of the semi-classical situation
nor constructed a Lagrangian, limiting ourselves to the presentation
of the Noether invariants defining the symplectic structure.
Therefore, the link between the abstract group parameters and the metric 
ingredients, needed to make contact with a gravitational theory, is less
obvious. In fact, the present model is more properly interpreted as a 
kind of {\it topological theory}, though with local degrees
of freedom,  where a metric notion must be recovered by imposing a 
constraint (in the spirit of General relativity from BF 
theory \cite{Pl77,CDJ91}) or by adding
an extra structure.

We present here different situations framed in a gravitational setting
where the mathematical construction we have developed plays a significant role.

\pagebreak

\ni{\it Space-time with boundary conditions}

\medskip

Perhaps the most conservative application of the mathematical
formalism developed in the previous section takes place in the context
of standard General Relativity  when boundary conditions play a significant
role. The study of the solution space of the (classical) theory in these 
scenarios shows that, due to the presence of these conditions, some of the
diffeomorphisms are not generated by the constraints of the theory
\cite{RT74} and therefore the former are not gauge transformations. 
As a consequence,
they actually move the solution space and constitute a dynamical symmetry.
In fact, in some models like asymptotic Anti de Sitter solutions to $2+1$
gravity \cite{BH86,Ba99}, which can be treated as a 
Chern-Simmons theory, or some treatments of black hole entropy 
\cite{Ca97}
these diffeomorphism parametrize the phase space of the theory providing
the physical degrees of freedom. This identification 
of the
solution space is a crucial first step for the quantization of the 
theory.

Abstract models for diffeomorphism dynamics like the one presented above
could prove to be useful in situations where the presence of more general 
boundary conditions confers a dynamical content to other subgroups
of the diffeomorphism symmetry. One cannot avoid the impression that
the limits in the application of these kind of techniques to more general 
scenarios is more related to our poor technical skills beyond the Virasoro 
symmetry than to actual conceptual reasons.

\vskip 0.6cm

\ni{\it Hilbert Space of a Gravitational Theory}

\medskip

\ni In the construction of a dynamical system out of a symmetry at the 
(semi-)classical level, there is a fundamental difference between the GAQ and,
for instance, a more standard coadjoint orbits method. In fact, 
the former provides not only a symplectic structure on which dynamics 
can be described, but also determines the Hamiltonian function(s) which 
specifies the evolution of the system. In this sense, our {\it topological}
model is endowed with an intrinsic dynamics dictated by the vector fields in
the characteristic module ${\cal G}_\Theta$. 

Nevertheless, we need not the full structure of GAQ to formulate 
well-defined models. We can resort to the GAQ as a powerful method to obtain, 
out of a fundamental symmetry, a 
well-defined phase space on which dynamics are
introduced by suplementing some extra information.

In this spirit, we are going to look at the quantum representation space 
of the previous section as the Hilbert space for a genuine
quantum gravitational theory. In this sense, our mathematical
construction supplies the {\it kinematics} of such a Gravity model.
The relevance of this Hilbert space for a gravitational theory relies on the
fact that the operators $L^\dagger_\mu({\bf m})$ and $L_\mu({\bf m})$
essentially constitute a quantum realization of the vector fields with support
on a manifold $M$. Taking advantage of the fact that a classical metric field
is a symmetric $(0,2)$ tensor field characterized by its action on such 
vector fields, we
are introducing a quantum metric operator $\hat{g}$ by defining
its action on some appropiate states (''quantized vector fields'' on a 
manifold).
An appropriate set of vector fields is the one defining 
a {\it tetrad} basis $\{e_a\}$. Considering the components
of these tetrads in a coordinate basis and formally expanding them in 
plane-waves, we find,
\bea
e_a(x)=e_a^\mu(x)\partial_\mu=\sum_{\bf m}e_a^\mu({\bf m})e^{i{\bf m}_\rho
x^\rho}\partial_\mu=\sum_{\bf m}e_a^\mu({\bf m})L_\mu({\bf m}) 
\eea
where we have made used of the  notation introduced in subsection 2.2.
Then, we promote the classical field $L_\mu({\bf m})$ to a quantum operator
by using the representation constructed in section 3,
\bea
e_a\mapsto e^\mu_a({\bf 0}) \hat{L}^\dagger_\mu({\bf 0})+
\sum_{m_f>0}\left(e^\mu_a({\bf -m})
\hat{L}^\dagger_\mu({\bf m})+
e^\mu_a({\bf m})\hat{L}_\mu({\bf m})\right) \ \ .
\eea
Associating with each of this operators a canonical state, $|e_a\rangle\equiv
\hat{e}_a|0\rangle$, we construct a metric operator $\hat{g}$ by imposing the 
{\it quantum tetrad condition},
\bea
\langle e_a|\hat{g}|e_b\rangle=\eta_{ab} \label{tetrad} \ \ .
\eea
A {\it minimal} manner (in the sense that
more general approaches could be followed) of implementing this condition 
consists in restricting 
the domain of the operator $\hat{g}$ to {\it one-particle} states 
$L^\dagger_\mu({\bf m})|0\rangle$. Even more, we demand it to be 
hermitian ($\hat{g}^\dagger=\hat{g}$) and diagonal on the 
${\bf m}$ label of these states: $\hat{g}L^\dagger_\mu({\bf m})|0\rangle=
g^\nu_\mu({\bf m })L^\dagger_\nu({\bf m})|0\rangle$.
In order to make a comparison with the classical theory, it is better to 
introduce an
object covariant in the two space-time indices, $g_{\mu\nu}({\bf m})\equiv
\langle 0|L_\mu ({\bf m}) \hat{g} L_\nu({\bf m})|0\rangle$, whose explicit 
form is
\bea
g_{\mu\nu}({\bf m})=\frac{-C_\rho}{2}\left(g^\sigma_\mu({\bf m}
(m_\nu\delta^\rho_\sigma+m_\sigma\delta^\rho_\nu)+
g^\sigma_\nu({\bf m}
(m_\mu\delta^\rho_\sigma+m_\sigma\delta^\rho_\mu)\right) \ \ ,
\eea
where we have considered for simplicity  that the only 
non-vanishing central parameter in (\ref{central}) is $C_\rho$.
With these elements, condition (\ref{tetrad}) simply expresses:
\bea
\eta_{ab}=\sum_{\bf m}g_{\mu\nu}({\bf m})
e^\mu_a({\bf m}){e^\nu_b}^*({\bf m}) \label{tetradexpl}
\eea
which is the analogue in this simplified example to the standard 
space-time variables expression and
relates the quantum action of $\hat{g}$ with a metric notion. 

The connection with the classical theory can be achieved by reversing the
step $L_\mu({\bf m})\mapsto \hat{L}_\mu({\bf m})$, and taking the
parameters $g_{\mu\nu}({\bf m})$ as the Fourier components of the classical
metric field:
\bea
g_{\mu\nu}(x)=\sum_{\bf m}e^{im_\rho x^\rho}g_{\mu\nu}({\bf m}) \ \ . 
\eea

\ni As we said before, the previous discussion  describes the 
{\it kinematics} of the 
metric degrees of freedom, but offers no clue on the way in which the specific
$g^\mu_\nu({\bf m})$ (or $e^\mu_a({\bf m})$) should be chosen. This is 
a {\it dynamical} information
that will be added in this approach by imposing a constraint, external to the
group,  on the $g_{\mu\nu}({\bf m})$ parameters. The simplest one we can
imagine, having a straightforward link to classical physics, is the 
Fourier transform of Einstein equations in vacuum, $R_{\mu\nu}=0$.
We find,
\bea
0&=&m_\lambda\sum_{\bf m}g^{\lambda\sigma}({\bf m-n})\left(n_\nu
g_{\mu\sigma}({\bf n})+n_\mu g_{\nu\sigma}({\bf n})-n_\sigma
g_{\nu\mu}({\bf n})\right)- \nn \\
&-&n_\nu\sum_{\bf n}g^{\lambda\sigma}({\bf m-n})\left(n_\lambda
g_{\mu\sigma}({\bf n})+n_\mu g_{\lambda\sigma}({\bf n})-n_\sigma
g_{\nu\mu}({\bf n})\right)+\nn \\
&+&\sum_{\bf n,p,q}g^{\eta\sigma}({\bf m-n-p})\left(p_\nu g_{\mu\sigma}
({\bf p})+p_\mu g_{\nu\sigma}({\bf p})-p_\sigma g_{\nu\mu}({\bf p})\right) \\
&& \ \ \ \ g^{\lambda\sigma'}({\bf n-q})\left(q_\lambda g_{\eta\sigma'}
({\bf q})
+q_\eta g_{\lambda\sigma'}({\bf q})-q_{\sigma'} 
g_{\lambda\eta}({\bf q})\right)- \nn \\
&-&\sum_{\bf n,p,q}g^{\eta\sigma}({\bf m-n-p})\left(p_\lambda g_{\mu\sigma}
({\bf p})+p_\mu g_{\lambda\sigma}({\bf p})-p_\sigma g_{\lambda\mu}({\bf p})
\right) \nn \\
&& \ \ \ \
g^{\lambda\sigma'}({\bf n-q})\left(q_\nu g_{\eta\sigma'}({\bf q})
+q_\eta g_{\nu\sigma'}({\bf q})-q_{\sigma'} 
g_{\nu\eta}({\bf q})\right) \nn
\eea
A model where (a sector of) the quantum phase space is constructed out of the 
diffeomorphism modes, corresponds to a theory where (part of) the 
diffeomorphisms are not gauge transformations at the quantum level. We can
then interpret the states in the Hilbert space as physical distortions
of a reference space-time corresponding to the maximum weight vector in the
representation. 
Therefore, $|0\rangle$ would correspond to a locally flat space-time, where
{\it locally} refers here to an entire
neighbourhood, not only to a point. One-particle states, which  display a 
direct 
link with the {\it quantized} vector fields, would correspond to 
local excitations of the metric field in a given space-time direction,
creating a distortion in distances measured by the action of the quantum 
operator $\hat{g}$. Even tough the restricted definition of $\hat{g}$ to
only the linear subspace of one-particle states is enough to show 
quantum features
such as the probability transition between different metric excitation states,
one can think of endowing multi-particle states also with the metric content.
A simple way of extending the domain of $\hat{g}$ consists in defining 
a proyector $P$ $(P^2=P)$ from the total representation Hilbert space to that 
of mono-particular states. Then, we define
$\hat{g}|\Psi\rangle\equiv \hat{g}P|\Psi\rangle$, so that an arbitrary
state acquires a metric excitation interpretation. 

\vskip 0.6cm

\ni{\it Gauge Gravity Theories}

\medskip

\ni
In the GAQ, the group law of the starting symmetry comprises
both the {\it connection fields} parameters and the ones corresponding to
the local or gauge group itself. However, and similarly to the shown
electromagnetic
case, a subtle non-trivial mixing between these two sets
of parameters occurs which implies a restructuration of the gauge and physical
degrees of freedom. As a result of this process, the original elements of the
gauge group do gain a dynamical content. 
When approaching gravity by using local diffeomorphisms as the symmetry 
associated with local translations, the
$l^\mu({\bf m})$ parameters themselves acquire a metric meaning.
Of course, the complete theory needs the addition of the other part of the
group, something that could be faced in principle in different ways
(directly metric structures, tetrad fields, Ashtekar connection variables...).
But independently on the actual way in which this missing structure is 
introduced, combinations of $m_\mu\tilde{X}^R_{l^\nu({\bf m})}$
will enter the theory as part of the physical gravitational degrees of 
freedom. Apart from that, the presence of new fields like $S^\rho({\bf m})$ 
in the extended diffeomorphism group, opens the possibility of constructing 
the {\it matter} side of the Einstein-like equations out of the symmetry 
structure.
In the spirit of a gauge approach to Gravity, the construction 
in section 3 must be understood
as an indermediate step in the development of a yet-to-be-completed theory.
Regarding the breakdown of diffeomorphism invariance, the presence of central 
terms in the commutator of the diffeomorphism generators turn then into a
set of second-class contraints. 
Not all the {\it spurious} degrees of freedom can be eliminated, thus
appearing new physical modes. This is analogous to the 
emergence of Proca field in \cite{Manolo1} or to the mechanism for 
cohomological mass generation of gauge bosons in \cite{Manolo2}.
Finally, the secondary goal of
advancing in the understanding of the interaction mixing is accomplished
by making explicit the non-trivial coupling of the diffeomorphisms and the 
local $U(1)$ symmetry.
In a general sense beyond the specific GAQ, the insight into 
diffeomorphism representations provided in section 3, will help the 
understanding of the imposition of diffeomorphism constraints in 
diffeomorphism invariant theories where a gauge is not fixed prior to 
quantization.

\vskip 0.6cm

\ni{\it Phenomenological Theories}

\medskip

The last physical interpretation promotes our construction 
to an effective theory valid
at low-intermediate energies. An underlying
more fundamental theory is assumed to exist at higher energies, with the 
only constraint that
diffeomorphism invariance,  as a gauge symmetry, must play a fundamental role 
in its formulation. Lacking such a description, though candidates do exist, a
phenomenological one is  proposed leaning on two basic assumptions. On the one 
hand, the specific space-time
dynamics are decoupled and described by the (semi-)classical theory 
(General Relativity or, more generally, Quantum Field Theory on curved 
spaces) and, on the other hand, the diffeomorphism invariance 
is anomalously realized in the quantum process which describes 
the implementation of the matter fields. This 
anomalous realization gives
rise to effective deformations of the fundamental diffeomorphism 
algebra ($S^\mu({\bf m})$ 
field) and confers with a dynamical content to some of the, in principle 
spurious, gauge diffeomorphism degrees of freedom (the implementation
of the diffeomorphism gauge invariance breakdown is explicit here). 
In this sense, the 
occurrence of an anomaly reveals the presence of more fundamental underlying 
physics at higher energies whose degrees of freedom are encoded at these
intermediate energies in the emergent effective degrees of freedom
\cite{Preskill}. 
The particular model in this paper can be seen as a generalization to
higher dimensions of the example presented in \cite{masa}. 
This reference provides a more detailed description of the physics involved in
this approach and suggests a mechanism for 
particle mass generation. Identifying
$\hat{L}_\mu({\bf 0})$ with the Hamiltonian of the complete physical system, 
in a given coordinate
system, and looking at the explicit form of its associated Noether 
invariant in terms
of those ones corresponding to the physical degrees of freedom (see
(\ref{Noether})), we find the general form
\bea
\hat{H}_{eff}=\hat{H}_{free}\left(\hat{\Phi}^\dagger,\hat{\Phi}\right)+
\hat{H}_{mixing}\left(\hat{\Phi}^\dagger,\hat{\Phi},
(\hat{L}_{\mu})^\dagger,\hat{L}_{\mu},(\hat{F})^\dagger,
\hat{F}\right) \ \ , \label{hamiltoniano}
\eea
where $\hat{\Phi}$ denote the non-gravitational fundamental degrees of freedom
(here $\hat{A}$, which can be seen as a scalar field),
{$\hat{L}_{\mu}$ are the dynamical diffeomorphisms and $\hat{F}$
are the extra anomalous effective degrees of freedom ($\hat{S}^{\mu}$ 
in our case). As we can see, the presence of such an anomaly 
entails corrections to the spectrum of the fundamental non-gravitational
degrees of freedom, opening the above-mentioned possibility of relating 
the emergence of mass terms to them.
\ni Finally, we must point out that in contrast with the first physical
interpretation, this phenomenological approach takes full advantage of the
structures obtained with the GAQ. In fact, not only a representation 
space is constructed but also the dynamics is determined by the choice
of the Hamiltonian among the generators inside ${\cal G}_\Theta$.

\section{Conclusions}

The main result of the present work is the construction of a maximum-weight
representation corresponding to the infinite-dimensional Lie algebra closed 
by the local diffeomorphisms $L_\mu({\bf m})$ defined on a manifold of 
dimension superior to one, acting in a semi-direct way on scalar
modes $A({\bf m})$ and finally non-centrally extended by the 
generator $S^\rho({\bf m})$.
Although this kind of representations are already studied in \cite{Larsson2},
the singularity of ours is related to its explicit (non-formal)
character and to the central role played by the concept of 
pseudo-cohomology. On the other hand, it fulfills the generalization of the
mathematical structure in \cite{QG1199}.

The physical relevance of this construction in the setting of (quantum)
gravitational dynamics is exemplified by the discussion of several different
approaches to the general problem. They all share the 
feature of confering a priviledged role to the notion of symmetry
in the construction of the physical system.
Our main conclusion is that the diffeomorphism group by its own cannot 
account for the complete gravitational dynamics (the basic aim in 
\cite{QG1199}), since it basicly defines a kind of topological theory that 
needs to be suplemented with additional structure (either including 
it inside a bigger
group, or by imposing a constraint, etcetera) in order to 
make contact with metric notions. However, this algebra
proves to play a key role in the 
formulation of some non-trivial physical models for quantum gravity,
where the issues of quantum diffeomorphism invariance breakdown 
and the mixing with internal symmetries are raised.
They suggest, in fact, some potential experimental consequences such 
as the presence of extra fields 
corresponding to non-central extensions ($S^\rho$) or the corrections to 
non-gravitational fields propagators (phenomelogical model).

\section{Acknowledgements}
One of us, J.L.J. would like to thank the Mathematical Department of the 
Queen Mary and 
Westfield College (University of London), and especifically to Prof. S. Majid, 
for the hospitality during the development of part of the work.


\begin{thebibliography}{99}

\bibitem{QG1199}  V. Aldaya, J.L. Jaramillo, Class. Quant. Grav. {\bf 17},
                1649 (2000). \\
V. Aldaya, J.L. Jaramillo, Class. Quant. Grav. {\bf 17},
                4877 (2000).

\bibitem{Polyakov} A.M. Polyakov, Mod. Phys. Lett. {\bf A11}, 893 (1987).

\bibitem{GAQ} V. Aldaya, J. Navarro-Salas, A. Ram\'\i rez, Commun. Math. Phys.,
{\bf 121}, 541 (1989).

\bibitem{electrograv} V. Aldaya, J.L. Jaramillo and J. Guerrero, 
{\it to appear in J. Phys {\bf A}}, {\bf hep-th/}0105278. \\
V. Aldaya, J.L. Jaramillo and J. Guerrero, 
{\it submitted}.

\bibitem{Manolo1} V. Aldaya, M. Calixto, M. Navarro, Int.J.Mod.Phys. 
{\bf A12}, 3609 (1997).

\bibitem{Manolo2} M. Calixto, V. Aldaya, J. Phys. {\bf A 32}, 7287 (1999).

\bibitem{Kibble} T.W.B. Kibble, J. Math. Phys. {\bf 2}, 212 (1961).

\bibitem{Al89} E. \'Alvarez, Rev. Mod. Phys. {\bf 61}, 561 (1989). 

\bibitem{Woodhouse} N.M.J. Woodhouse, {\it Geometric Quantization}, Oxford
University Press, (1991).

\bibitem{Fuks} D.B. Fuks, {\it Cohomology of infinite dimensional
algebras}, Consultants Bureau (Russian ed. (1984)), 1986

\bibitem{pseudoext} V. Aldaya, J. Guerrero, Rep. Math. Phys. {\bf 47}, 213
(2001).

\bibitem{Larsson} T.A. Larsson, {\bf math-ph/}0002016.

\bibitem{Dhu}A. Dzhumadil'daev, Z. Phys. C {\bf 72}, 509 (1996).


\bibitem{Utiyama} R. Utiyama, Phys. Rev. {\bf 101}, 1597 (1956). 

\bibitem{Serre} J.P. Serre, {\it Lie algebras and Lie groups}, New Yor, 
Benjamin (1965).


\bibitem{Pepe} V. Aldaya, J. Navarro-Salas, Commun. Math. Phys. 
                        {\bf 113}, 375 (1987).

\bibitem{Saletan}E.J. Saletan, J. Math. Phys. {\bf 2}, 1 (1961).

\bibitem{Inonu-Poin} V. Aldaya, J. A. de Azc\'arraga, Int. J. Theo. Phys.
{\bf 24}, 141 (1985).

\bibitem{Pepe2} V. Aldaya, J. Navarro-Salas, Commun. Math. Phys. {\bf 113}
375 (1987).

\bibitem{Pl77} J. Plebanski, J. Math. Phys. {\bf 18}, 2511 (1977).

\bibitem{CDJ91} R. Capovilla, J. Dell, T. Jacobson, Class. Quant. Grav
{\bf 8}, 59 (1991). 

\bibitem{RT74} T. Regge, C. Teitelboim, Ann. Phys. (N.Y.) {\bf 88},
286 (1974).

\bibitem{BH86} J.D. Brown, M. Henneaux, Commun. Math. Phys. {\bf 104},
207 (1986).

\bibitem{Ba99} M. Ba\~nados,  Charla invitada en {\it Second Meeting 
"Trends in Theoretical Physics"}, Buenos Aires (1998); {\bf hep-th}9901148.

\bibitem{Ca97} S. Carlip, Nucl. Phys. Proc. Suppl. {\bf 57}, 8 (1997).


\bibitem{Preskill} J. Preskill, Ann. Phys.(N.Y.) {\bf 210}, 323 (1991).

\bibitem{masa}J.L. Jaramillo, V. Aldaya, Mod. Phys. Lett.  {\bf A17}, 809
(2002).

\bibitem{Larsson2} T.A. Larsson, Commun. Math. Phys. {\bf 201} (1999), 461.




\end{thebibliography}
\end{document}